\documentstyle[aps,prd,epsfig,floats,manuscript]{revtex}

\newcommand{\beq}[1]{\begin{equation}\label{#1}} 
\newcommand{\eeq}{\end{equation}} 

\author{Ph.~H\"agler$^1$, R. Kirschner$^2$, A. Sch\"afer$^1$
\footnote{e-mail: andreas1.schaefer@physik.uni-regensburg.de}, 
L.~Szymanowski$^1$\footnote{On leave of absence from Soltan 
Institute for Nuclear Studies, Warsaw},}               

\title{The gluon field of a fast moving nucleus and the
 effective langrangian for QCD at high energy } 

\begin{document}

\maketitle 
 
\begin{center} 
$^1$
Institut f\"ur Theoretische Physik\\
Universit\"at Regensburg\\
D-93040 Regensburg, Germany\\

\vskip.1in 
$^2$
Institut f\"ur Theoretische Physik\\
Universit\"at Leipzig\\
D-04109 Leipzig, Germany\\

\end{center}
\vskip 0.3cm

PCAS: 12.38.Bx, 12.38.Aw, 24.85+p

\vskip 0.3cm
\noindent{\bf Abstract:}   
\vskip 0.3 cm
Starting from the effective lagrangian 
for QCD at high energy   we calculate the lowest
perturbative contributions to the potential 
of  a relativistic nucleus and compare our results to those
derived by Kovchegov (see Y.V. Kovchegov, Phys. Rev. {\bf D55}, 5445 (1997)).  
The results differ already at order $g^3$ which can be traced to the fact
that the meaning of the underlying gluon fields is different.
(The effective gluon field we use is a gauge invariant object.)
Both approaches should therefore be seen as alternatives, the relative 
merits of which have to be judged by their phenomenological success.


\eject 
\newpage 
\section {Introduction}

In  Ref.\cite{McLV} McLerran and Venugopalan have proposed a program of
computing the gluon distribution function for a very large nucleus at small
values of the Bjorken variable $x$. In this approach the ultrarelativistic
nucleus looks like a pancake in the transverse directions and it is described by a 
classical colour potential whose form is characteristic for the shock wave
picture of high energy scattering. It was argued that
although the colour field of  each individual nucleon is so small that
perturbation theory can be applied, the total field is  strong enough to
justify  the classical approach. 
The explicit form of
this classical potential was found and studied subsequently by 
Kovchegov \cite{KI}, \cite{KII} in a 
special  model describing  a nucleus  as a set of nucleons 
each of which is a  colour
singlet dipole  built  of a  quark and an antiquark. 
He has shown also that this non-abelian Weizs\"acker-Williams potential
 leads to the same correlation functions for the gluon
distribution function as  the model by McLerran 
and Venugopalan \cite{KI}.  
We present an alternative approach
based on the effective lagrangian for QCD at high energy 
\cite{EL}, \cite{KLS}. 
As a first step in this direction we study in this contribution the
colour potential of an ultrarelativistic nucleus which plays the analogous
role as the classical WW-potential in the approach developed in Ref.\cite{McLV}.
 

The classical potential of a relativistic nucleus as derived 
 in \cite{KI} is  the sum of the potentials of  
the individual relativistic 
nucleons 
transformed to 
the light-cone gauge $A_- = A_0 - A_3 =0$. It has the following form \cite{KI}
\begin{eqnarray}
A^\rho (x^\bot,x_+)&=&\frac{g}{2\pi}\sum_{i=1}^N\bigg(S(x^\bot,x_{i+})(t^a_i)t^a 
S^{-1}(x^\bot,x_{i+})\frac{x^\rho-x_i^\rho}{|x^\bot-x_i^\bot|^2}\theta(x_+-x_{i+})
\nonumber\\
&&-S(x^\bot,x_{i+}')(t^a_i) t^a S^{-1}(x^\bot,x_{i+}')
\frac{x^\rho-x_i'^\rho}{|x^\bot-x_i'^\bot|^2}\theta(x_+-x_{i+}')\bigg),
\label{A}
\end{eqnarray}
where the transformation matrix $S(x)$ is given by
\begin{eqnarray}
S(x^\bot,x_+)\!\!&=&\!\! P\exp\bigg(\!\!-ig\int_{-\infty}^{x_+}d\tilde{x}_+
\bigg\{\frac{-g}{2\pi}\sum_{i=1}^N t^a(t^a_i)\big[\delta(\tilde{x}_+-x_{i+})
\ln(\lambda|x^\bot-x_i^\bot|)\nonumber\\
&&-\delta(\tilde{x}_+-x_{i+}')\ln(\lambda|x^\bot-x_i'^\bot|)\big]\bigg\}\bigg).
\label{S}
\end{eqnarray}
The vectors $x_i,x_i'$ are the positions of the quark and 
antiquark, respectively, in the i-th nucleon, and $\lambda$ is an infrared cutoff.  
Because the nucleons live in  seperated colour-spaces, 
the 'matrix-products' 
like $(t^a_1)(t^b_2)(t^c_1)$ are understood as $(t^a_1 t^c_1)_{i_1
j_1}(t^b_2)_{i_2 j_2}$, where 
the $t^a_i$ are the generators in the fundamental 
reprensentation of SU(3) acting in the colour space of the $i$-th nucleon. 
We use the following notation for the light-cone 
coordinates: $x_-\equiv x_0-x_3=x^0+x^3\equiv x^+$, $\partial_- x_+ =
\partial_+ x_- =1$. Throughout the paper the index $\rho$
describes the transverse components and always 
runs from $1$ to $2$. For a given ordering of positions in the variable $x_+$ of
the $N$ nucleons constituting the nucleus, e.g. for the ordering
$x_{N+} > x_{N-1\:+} > \dots x_{2+} > x_{1+}$, 
 the transformation matrix $S$ can be written in the following form \cite{KI}
\begin{equation}
\label{Sord}
  S(x)=\prod_{i=1}^N \exp\bigg(\frac{ig^2}{2\pi} t^a(t^a_i) 
\ln\bigg[\frac{|x^\bot-x^\bot_i|}{|x^\bot-x'^\bot_i|}\bigg]\theta(x_+-x_{i+})\bigg).
\end{equation}

The quantum structure of the Weizs\"acker-Williams field (\ref{A}) was 
studied  by Kovchegov in Ref.\cite{KII}. By expanding  eq.(\ref{A}) and
eq.(\ref{Sord}) in a power series in the coupling constant $g$ it was
shown that for two nucleons the terms of
eq.(\ref{A}) up
to order $g^5$  can be reproduced by calculating  
the corresponding  Feynman diagrams in the 
light-cone gauge $A_-=0$, if some specific assumptions are made. 
It was necessary to adapt a somewhat peculiar regularization prescription for 
the spurious pole in the gluon propagator, namely it was assumed
that the gluon propagator
 has the form  
\begin{equation}
P_{\mu\nu}(k) = -\frac{i}{k^2}\left(g_{\mu\nu} - \frac{\eta_\mu k_\nu}{k_- -
i\epsilon} - \frac{\eta_\nu k_\mu}{k_- + i\epsilon} \right) \;,
\end{equation}
where colour indices are suppressed and $\eta$ is the light-cone vector 
$\eta\cdot k = k_-$.


Before presenting our calculations let us remind briefly of some
basic properties of the effective lagrangian. 
The effective lagrangian approach for
 QCD
at high-energies  was proposed by Lipatov in \cite{EL}. In Refs.\cite{KLS} 
the effective
lagrangian determining the tree amplitudes for scattering 
in the leading power of the
scattering energy was derived from the original QCD lagrangian. 
This effective lagrangian was subsequently generalized to include the
next-to-leading logarithmic corrections \cite{ELn} and the next-to-leading
power corrections in the scattering energy \cite{KLn}. For the purpose of the present paper
it is however sufficient  to
consider only the effective lagrangian derived in \cite{KLS}. It is
expressed in terms of two types of the effective fields: 

a) $s$-channel fields which are almost on mass-shell and which describe 
 physical degrees of freedom of the
scattered and produced particles
propagating in the $s$-channel 

b)  $t$-channel fields of Coulomb type which are   
responsible for the transfer of the interactions and 
propagate in the $t$-channel. 
 
Also this effective lagrangian involves three types of interaction vertices:

a) the  triple scattering vertices describing the interaction 
of two $s$-channel fields and one $t$-channel Coulomb field,

b) the triple production vertices describing the production of one
$s$-channel particle out of two $t$-channel Coulomb fields,

c) the triple vertices describing the interaction of three $t$-channel Coulomb fields.

We want to emphasize that the t-channel fields are gauge invariant objects. 
This is one of the main differences between the calculation of the 
gauge-variant WW-potential by Kovchegov and our following analysis.
Closer analysis of the calculations  done in  \cite{KII} 
for the classical potential (\ref{A})
and in particular a detailed  comparison of the contributing Feynman diagrams 
with the structure of the effective lagrangian derived in
\cite{KLS} 
suggests that there should be some relation between both approaches which we
would like to investigate. 
As the effective lagrangian is based on the summation of 
tree-level amplitudes in the Regge kinematics 
its predictions should be close to those
of quasi-classical approaches. Therefore a comparison with
Kovchegov's results is meaningful.
As a first step in this direction we performed 
similar calculations up to order $g^5$ 
using the effective lagrangian.
 To allow for a direct comparison we use the same
regularization conventions for  the spurious poles as in \cite{KII}.

\section{The colour potential of the nucleus}

From the whole set of interaction vertices of the effective lagrangian given in \cite{KLS}
    calculations of the potential described by eq.(\ref{A}) involve only
 the scattering vertices describing the interaction of a
quark current with large momentum component $p_-$ with a $t$-channel Coulomb field
$A_+^a$ 
and the interaction vertex of two $t$-channel Coulomb fields $A_-^a$ with
one Coulomb field $A_+^a$
\begin{equation}
\label{int}
{\cal L}_{int}=\frac{g}{2}\overline\Psi_-\gamma_-t^a\Psi_-A_+^a
-\frac{gf^{abc}}{8}\bigg(\frac{1}{\partial_-}A_-^b\bigg)A_-^c(\partial_\rho\partial^\rho A_+^a)
\end{equation}
where the meaning of the operator $\frac{1}{\partial_-}$ is discussed below.
The corresponding kinetic
terms for the $s$-channel fermions and Coulomb fields are
\begin{equation}
\label{kin} 
{\cal L}_{kin}=i\overline\Psi\!\not{\!\partial}\Psi+\frac{1}{2}
A_+^a\partial_\rho\partial^\rho A_-^a
\end{equation}
The fields $A^a_+$ and $A^a_-$ in eqs. (\ref{int}), (\ref{kin}) describing
the gluonic reggeons are related to the $t$-channel modes of the original
transverse gluon potential $A^{a\rho}$
\begin{equation}
\label{Apm}
A^a_+ = -\frac{1}{\partial_-}\partial^\rho A^a_\rho \;, \;\;\; 
A^a_- = -2 \frac{\partial_- \partial_\sigma}{\partial_\rho \partial^\rho}
A^{a \sigma}
\end{equation}
(we remind that the indices $\sigma$, $\rho$ describe 
the transverse components and take
the values $\rho, \sigma = 1,2$ ).
This has been obtained in the axial gauge with the minus component of the
original gluon potential set to zero, $A_-=0$, and integrating out over
the plus component of the original gluon potential $A_+$. 
Let us emphasize that the form of the kinetic term for the $A_\pm$ fields 
(\ref{kin}) and
the interaction vertex (\ref{int}) 
reflect an important property of the underlying  multi-Regge kinematics:
The $k_-$ momenta of Coulomb fields in the $t$-channel are strongly ordered
and decrease in our case from the left to the right.
In order to compare the results of our calculations with those of
Ref.\cite{KII} we have also to introduce the coupling of the $A^a_-$ field
to an external current with transverse components $J^a_\rho$. 
Taking into account the form of the first term in the interaction lagrangian
(\ref{int}) and the 
definition (\ref{Apm})
 of the $A_+^a$ field  
  we use the following form of the coupling 
\begin{equation}
\label{J}
{\cal L}_{J}= \frac{1}{2}A_-^a J_+^a 
\;.
\end{equation}
The sum 
\begin{equation}
{\cal L}={\cal L}_{kin}+{\cal L}_{int}+{\cal L}_{J}
\label{sum}
\end{equation}
of the expressions (\ref{kin}), (\ref{int}) and (\ref{J})
 defines the
part of the effective lagrangian relevant for our calculations.   

We have still to define  the meaning of the operator $\frac{1}{\partial_-}$ in
eqs. (\ref{int}) and (\ref{J}). Unfortunately, the  derivation of
the effective lagrangian given in \cite{KLS} does not fix this
operator unambigously and we have to make an additional assumption.
We understand that the operator $\frac{1}{\partial_-}$ is regularized
according to the Mandelstam-Leibbrandt-scheme \cite{ML} which in our opinion
has a sound theoretical foundation.
It turns out that this prescription in the case of our calculation of the
potential (\ref{A}) is equivalent to the prescription of the contributing
terms in \cite{KII} (see discussion below).



Now we are in the position to write down the Feynman rules for our effective
lagrangian. They are summarized in  Fig 1.
%
%
%
%



\begin{figure}[h]
\begin{minipage}{50mm}
\begin{center}
\epsfig{file=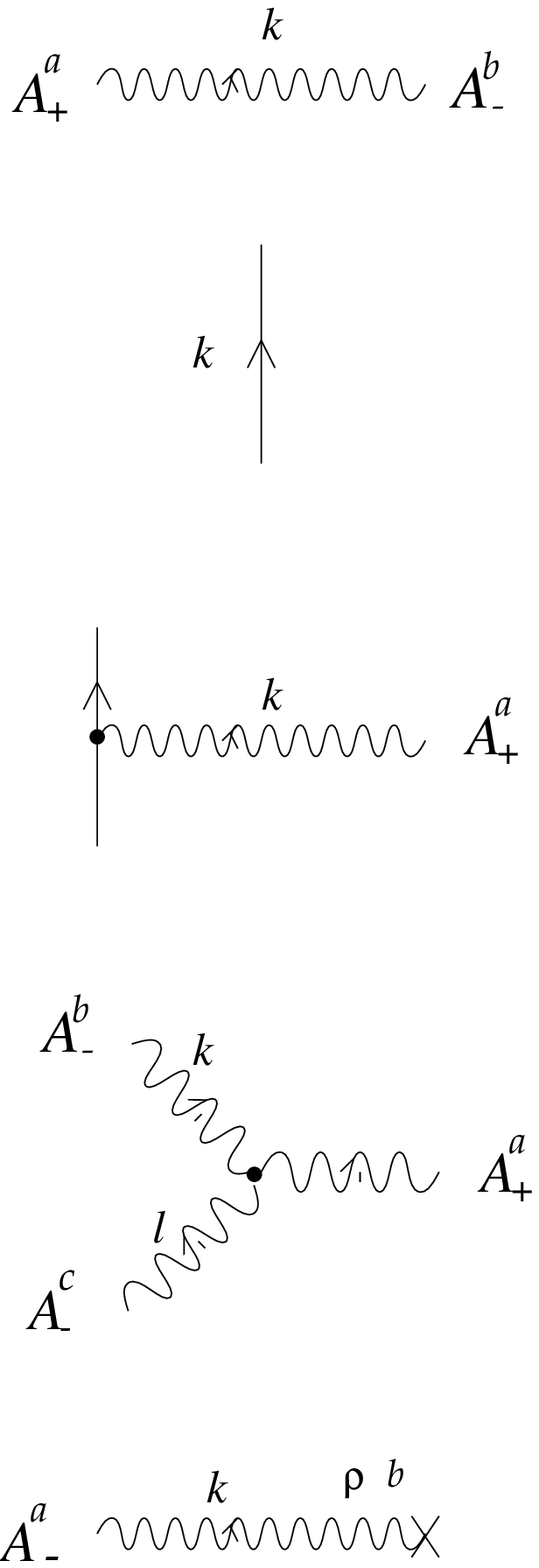, height=8cm,width=3.5cm}
\end{center}
\end{minipage}
\hspace{1cm}
\begin{minipage}{50mm}
\begin{eqnarray}
\label{frule1}
&-2i/(k^\rho k_\rho)&\nonumber
\\[0.5cm] 
&i/\!\!\not{\! p}&\nonumber
\\
\nonumber
&&\\
&\frac{1}{2}ig\gamma_-t^a&\nonumber
\\[0.8cm]
&\frac{1}{4}gf^{abc}\bigg(1/(k_-+ i\epsilon{\textrm sgn}(k_+))&
\nonumber\\
&-1/(l_-+ i\epsilon{\textrm sgn}(l_+))\bigg)(k^\bot+l^\bot)^2&\,\,\,\,\,\,
\nonumber\\[0.4cm]
\label{frule5}
&-ik^\rho/(k_- +i\epsilon {\textrm sgn}(k_+))&\nonumber
\end{eqnarray}
\end{minipage}
\caption{Feynman rules corresponding to the effective lagrangian
(\ref{sum}).}
\end{figure}
\hfill
\hfill

Using these Feynman rules we can calculate the
contribution of order $g$
to the ${\cal T}$-matrix element
(${\cal S}=1 +i{\cal T}$) corresponding to the diagram in Fig.2.

\begin{figure}[h]
\begin{center}
\epsfig{file=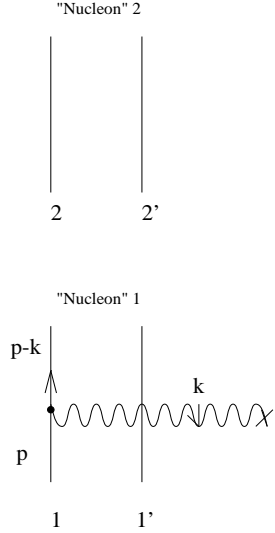, height=7cm,width=3.5cm}
\caption{First order graph.}
\end{center}
\end{figure}

Demanding that  the quark in the final state is on mass-shell we
multiply the resulting expression by $2\pi \delta((p-k)^2)$ 
and obtain the order
$g$ contribution to the potential in the momentum space
\begin{equation}
\label{gm}
A^\rho_1(k) = 2gt^a(t^a_1) \frac{k^\rho}{(k^\bot)^2}\frac{1}{k_--i\epsilon}2\pi\delta(k_+)
\end{equation}

Note that the product of the pole term in $k_-$ and $\delta(k_+)$ appears as
the limit of the Mandelstam-Leibbrandt prescription
\begin{equation}
\frac{1}{k_- + i\epsilon {\textrm sgn(k_+)}}\delta\bigg(k_+(1-\frac{k_-}{p_-})) + 
\frac{k^{\perp 2}}{p_-}\bigg) \stackrel{{\textrm for} \,\,p_- \to \infty\,}{\longrightarrow}
\frac{1}{k_- - i\epsilon}\delta(k_+)\;
\end{equation}
and it coincides with the regularization used in \cite{KII}.

Expression (\ref{gm}) of course agrees with the corresponding
contribution in \cite{KII} (see eq.(5)), so after
 performing  a Fourier transformation 
\begin{equation}
A^\rho_1(x)
 = \int \frac{d^4k}{(2\pi)^4} A^\rho_1(k) e^{ik(x-x_1)}
\end{equation}
 and taking into account
  the interaction with the other lines shown in Fig.2 we reproduce the
 lowest order contribution to the potential (\ref{A}),
 $A(x^\perp ,x_+)|_g$:
\begin{eqnarray}
\nonumber
A^\rho(x^\perp,x_+)|_{g\, eff.lagr.} &=& A^\rho(x^\perp ,x_+)|_g =\\ \nonumber
&=&\frac{g}{2\pi}t^a\bigg(\frac{x^\rho-x_1^\rho}{|x^\bot-x_1^\bot|^2}\theta(x_+-x_{1+})(t^a_1)
+\frac{x^\rho-x_2^\rho}{|x^\bot-x_2^\bot|^2}\theta(x_+-x_{2+})(t^a_2)\bigg)\\
&&-\frac{g}{2\pi}t^a\bigg(\frac{x^\rho-x_1'^\rho}{|x^\bot-x_1'^\bot|^2}
\theta(x_+-x_{1+}')(t^a_1)+\frac{x^\rho-x_2'^\rho}{|x^\bot-x_2'^\bot|^2}
\theta(x_+-x_{2+}')(t^a_2)\bigg)
\label{g} 
\end{eqnarray}

Passing to the $g^3$-contribution let us note that contrary to the
calculations done in \cite{KII}, in the case of the
effective lagrangian approach we have to calculate only
one diagram shown in Fig.3. 

\begin{figure}[h]
\begin{center}
\epsfig{file=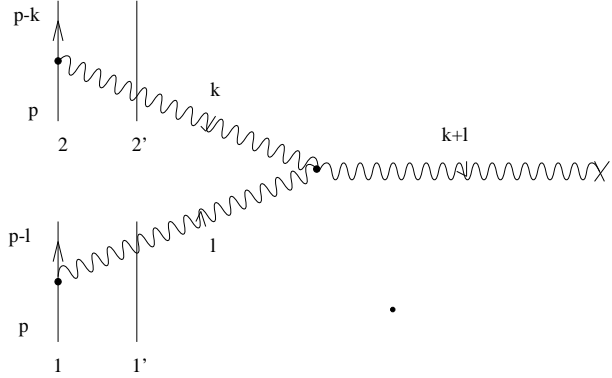, height=5cm,width=8cm}
\caption{Third order graph.}
\end{center}
\label{kfig3}
\end{figure}

Its contribution to the ${\cal T}$-matrix
multiplied by the
corresponding factors and putting two lines in the final state
on the mass-shell has the form
\begin{eqnarray}
A^\rho_3(k,l) &=&-2ig^3t^cf^{bac}(t^a_2)(t^b_1)\frac{(k+l)^\rho}{(l^\bot)^2(k^\bot)^2}
\frac{1}{(k+l)_--i\epsilon}
\bigg(\frac{1}{l_--i\epsilon}-\frac{1}{k_--i\epsilon}\bigg)(2\pi)^2\delta(k_+)\delta(l_+)
\label{g3m}
\end{eqnarray}
which has a similar but not identical structure as the final result given in
\cite{KII} in eq.(10). By perfoming a double Fourier transform
of expression (\ref{g3m})
\begin{equation}
\nonumber
A^\rho_3(x^\perp,x_-)= \int \frac{d^4k}{(2\pi)^4} \frac{d^4l}{(2\pi)^4}
A^\rho_3(k,l) e^{il(x-x_1) +ik(x-x_2)}
\end{equation}
and imposing the condition
$x_{2+}> x_{1+}$ we see that only  the term with the pole in $l_-$ gives a
contribution for such ordering. Next, taking into account the remaining
three contributions corresponding to the interaction of the gluons $k$ and $l$
with the lines $1\;2'$, $1'\;2$ and $1'\;2'$ we obtain the formula
\begin{eqnarray}
\nonumber
A^\rho(x^\perp,x_-)|_{g^3, eff.lagr.} &=&-\frac{g^3}{2(2\pi)^2}t^a f^{abc}(t^b_1)(t^c_2)
\bigg\{\theta(x_+-x_{2+})\partial^\rho\bigg[\ln(\lambda|x^\bot-x_2^\bot|)
\ln\frac{|x^\bot-x_1^\bot|}{|x^\bot-x'^\bot_1|}\bigg]\nonumber\\
&&-\theta(x_+-x_{2+}')
\partial^\rho\bigg[\ln(\lambda|x^\bot-x_2'^\bot|)
\ln\frac{|x^\bot-x_1^\bot|}{|x^\bot-x'^\bot_1|}\bigg]\bigg\}\;,
\label{g3el}
\end{eqnarray}
where the transverse gradient $\partial^\rho$ always acts on $x^\perp$.
This result has to be compared with the $g^3$ contribution to the potential
(\ref{A})
\begin{eqnarray}
A^\rho(x^\perp,x_-)|_{g^3} \!\!&=&\!\!-\frac{g^3}{(2\pi)^2}t^a f^{abc}(t^b_1)(t^c_2)
\ln\bigg[\frac{|x^\bot-x^\bot_1|}{|x^\bot-x'^\bot_1|}\bigg]
\bigg(\frac{x^\rho-x_2^\rho}{|x^\bot-x_2^\bot|^2} \theta(x_+-x_{2+})\nonumber\\
\!\!& &-\frac{x^\rho-x_2'^\rho}{|x^\bot-x_2'^\bot|^2} \theta(x_+-x_{2+}')\bigg).
\label{g3}
\end{eqnarray}
We see that although both expressions (\ref{g3el}) and (\ref{g3}) have a
similar structure they differ substantially. The main difference apart of an overall
coefficient lies in the fact that in the effective lagragian result
the transverse gradient acts on both logarithms (which correspond to the
propagators of the gluons with momenta $k$ and $l$), whereas
in eq.(\ref{g3}) it acts only on one of them. 
Let us remark that the $\lambda$-dependence of eq. (\ref{g3el}) drops
out in the limit $x_+>>x_{2+},x'_{2+}$.

In order to calculate the contribution of order $g^5$ averaged in
the colour space of nucleon 1 in our effective theory 
we have to take into account 
only two diagrams shown in Fig. 4 (the analogous calculations of Feynman
diagrams in
the light-cone gauge involve 13 diagrams).
\begin{figure}[h]
\begin{center}
\epsfig{file=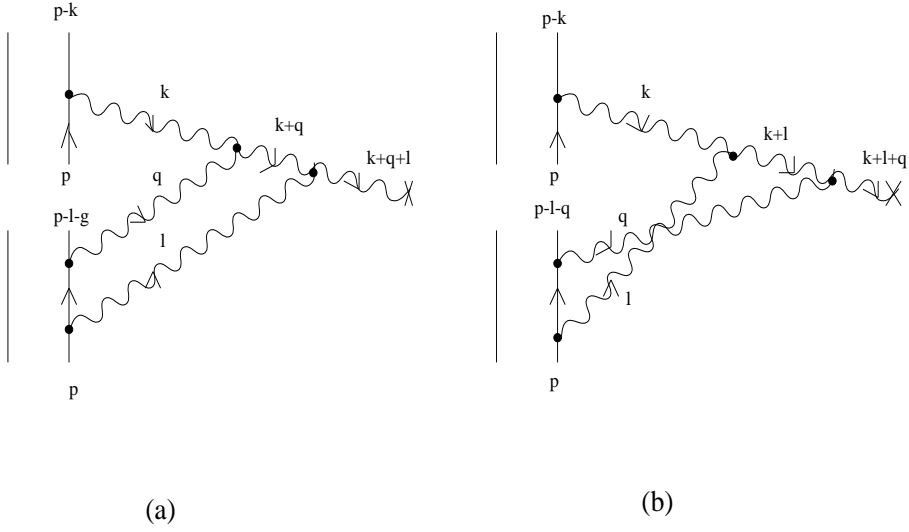, height=7cm,width=12cm}
\caption{Fifth order graphs.}
\end{center}
\label{kfig4}
\end{figure}


Calculating these diagrams we introduce two delta function factors 
which put two external lines in the final state on mass shell. 
Let us note that we are only interested in the classical contribution 
of the diagrams and therefore do not perform the loop-momentum 
integration but symmetrize the contributions with respect 
to $l$ and $q$ and use the following substitution

\begin{equation} 
\label{subst}
\frac{1}{l_+ - i\epsilon} \delta(q_+ + l_+) \to 
\frac{1}{2}\left( \frac{1}{l_+ - i\epsilon} + \frac{1}{q_+ - i\epsilon}
\right)\delta(q_+ + l_+) = i\pi \delta(l_+)\delta(q_+)\;.
\end{equation}

The resulting expression is 

\begin{eqnarray}
<A^\rho_5(k,l,q)>_1 &=&\!g^5t^a(t_2^a)\frac{(k+l+q)^\rho}{(k^\bot)^2(l^\bot)^2(q^\bot)^2}
\frac{1}{(k+l+q)_--i\epsilon}\nonumber\\
&&\!\times\bigg(\frac{1}{(q_--i\epsilon)(l_--i\epsilon)}
-\frac{1}{(q_--i\epsilon)((k+q)_--i\epsilon)}\nonumber\\
&&\!-\frac{1}{(l_--i\epsilon)((k+l)_--i\epsilon)}\bigg)(2\pi)^3\delta(k_+)\delta(l_+)\delta(q_+).
\label{g5m}
\end{eqnarray} 

We obtain the same answer if we put directly the fermionic line with
momentum $p-l$ in the nucleon 1 on the mass-shell, i.e. if we take 
only the contribution of the delta function $i\pi\delta(l_+)$ from the propagator into account.



By calculating now the triple Fourier transform of (\ref{g5m})
\[
<A^\rho_5(x^\perp,x_-)>_1 = \int
\frac{d^4k}{(2\pi)^4}\frac{d^4l}{(2\pi)^4}\frac{d^4q}{(2\pi)^4}
<A^\rho_5(k,l,q)>_1 
e^{il(x-x_1) +iq(x-x_1) + ik(x-x_2)}
\]
we obtain the expression
\begin{eqnarray}
<A^\rho_5(x^\perp,x_-)>_1|_{eff.lagr.} &=&-\frac{g^5}{8(2\pi)^3}t^a(t_2^a)\theta(x_+-x_{2+})
\partial^\rho\bigg(\ln^2(\lambda|x^\bot-x_1^\bot|)\ln(\lambda|x^\bot-x_2^\bot|)\bigg)
\label{A5x}
\end{eqnarray}
which we can compare with the corresponding term obtained from the expansion
of the classical potential (\ref{A}) (see eq.(15) in \cite{KII}),  being
reproduced by the sum of the Feynman diagrams shown in Fig.5 in \cite{KII}
\begin{equation}
\label{A5K}
<A^\rho_5(x^\perp,x_-)>_1 =-\frac{g^5}{4(2\pi)^3}t^a(t_2^a)\theta(x_+-x_{2+})
\ln^2(\lambda|x^\bot-x_1^\bot|)\frac{x^\rho-x_2^\rho}{|x^\bot-x_2^\bot|^2}.
\end{equation}
Again we see that the main difference between both expressions (apart
from a numerical factor of 2) is related to
the fact that in (\ref{A5x}) the transverse gradient acts on all propagators
of the gluons with momenta $k$, $l$ and $q$ whereas in (\ref{A5K}) it acts
only on the propagator of the gluon with momentum $k$. 
Let us note that this feature of our result will persist in higher orders.
The reason for this lies in the factorized form of the vertices in the 
longitudinal and transverse parts and in the fact that the transverse gradient
 is directly related to the coupling of the external current $J^a_\rho$. 
As a consequence it will always act on the whole expression of each diagram.

\section{Discussion}

On the basis of our calculations we  conclude that the potential
obtained with the use of the effective
lagrangian for high energy QCD differs from the classical potential of
the  ultrarelativistic nucleus  on which the approach developed in 
\cite{McLV} is based. This fact by itself should not be very surprising since
as we already mentioned the classical and the effective
potential correspond to different types of fields (our effective Gluon
is a Reggeon).
Taking into account that in both methods only classical contributions are 
calculated, it is suprising that the differences appear already in the
lowest orders of perturbation theory. We expect these differences to 
persist in calculations for physical observables. As the effective
lagrangian is derived from QCD with controlled, systematic and relatively
mild assumptions we hope that our method will prove to be more
successful in phenomenological applications.
Presently our approach and that of Ref.\cite{McLV} should simply be seen 
as alternatives.
We want
to stress also that our calculations along the line of \cite{KII} 
for the $g^5$-terms are very efficient because they build on 
the work invested in the derivation of the effective lagrangian.


\newpage
{\bf\Large Acknowledgements}

\vskip.2in
We acknowledge stimulating discussions with Y. Kovchegov about 
the connection between both approaches.

L.Sz. would like to thank Lev Lipatov for discussion.

This work was supported by GSI and DFG. R.K. and L.Sz. would like to
acknowledge the support from the German-Polish agreement on 
scientific and technological cooperation N-115-95.

\end{document}